\documentclass{article}
\usepackage{epsfig}
\usepackage{amssymb}
\usepackage{amsmath}
\usepackage{amsfonts}
\newcommand \be{\begin{eqnarray}}
\newcommand \ee{\end{eqnarray}}
\begin{document}
\begin{center}
{\bf Harmonic Oscillator trap  and the phase-shift approximation }\\
\bigskip
\bigskip
H. S. K\"ohler \footnote{e-mail: kohlers@u.arizona.edu} \\
{\em Physics Department, University of Arizona, Tucson, Arizona
85721,USA}\\
\end{center}
\date{\today}

\begin{abstract}
The energy-spectrum of two point-like particles 
interacting in a 3-D isotropic Harmonic Oscillator (H.O.) trap is related
to the free scattering phase-shifts $\delta$ of the particles by a
formula first published by Busch et al.
It is here used to find an expression for  the \it shift \rm
of the energy levels, caused by the interaction,
rather than the perturbed spectrum itself.
In the limit of high energy (large quantum number $n$ of the H.O.) 
this shift is shown to be given  by $-2\frac{\delta}{\pi}$,
also valid  in the limit of infinite  as well as zero
scattering length at all H.O. energies.

Numerical investigation shows 
that the shifts  differ from the exact result of Busch et al,
by  less than $<\frac{1}{2}\%$ except for $n=0$ when it can be as large
as $\approx 2.5\%$.

This approximation for the energy-shift  is well known from another
exactly solvable model,
namely that of two particles interacting  in a spherical infinite 
square-well trap  (or box) of radius $R$ in the limit $R\rightarrow \infty$, 
and/or in the limit of large energy. It is in this context referred to as 
the \it phase-shift approximation \rm.

It can  be (and has been)  used   in  (infinite) nuclear matter 
calculations to calculate the two-body effective interaction
in situations where in-medium effects can be neglected.
It has also been used in expressing the energy of free electrons in a
metal.
\end{abstract}

\section{Introduction}
Recent years has seen a tremendous progress in the ability to study the
physics of atoms  trapped in confined optical lattices. A significant
development has been  the tuning of the interactions (scattering
lengths) between the particles via Feshbach resonances.\cite{blo08} 
A further important 
achievement is the ability to confine just a few atoms in a potential well 
or trap.\cite{koh05} The
situation of a few particles confined in this fashion is also encountered in
other areas of physics. One old problem in nuclear physics is for example 
that of two or more nucleons 'outside' a closed shell. The recent
experimental research in atomic physics that enables a control of 
interacions as well as the
environment represented by the trap has stimulated  theoretical
studies of this basic quantum-mechanical problem: two or more particles
interacting while trapped in some specified potential well. 

In this paper \footnote{A preliminary version of this report 
is found in ref. \cite{hsk11}} the investigation is restricted to 
two particles in either a 3-dimensional H.O. trap or in a spherical 
square-well trap. Of particular interest in a study of this kind is 
of course the two-body interaction between the particles. It can be 
represented by some potential model e.g. psudo-potential
\cite{blo02,rot13} or chiral effective field theory
potential\cite{yan13}. 

It is however found that the interactions, under specific asumptions,
can be related to scattering data and this is the route taken in
this report.

\subsection{3-D H.O. well}
An expression first derived by Busch et al \cite{bus98} relates the 
\it energy-spectrum \rm of  two  point-like particles, interacting 
in a 3-D Harmonic Oscillator (H.O.) well, to  scattering phase-shifts.  
The derived formula 
is restricted to point-like particles, meaning that the 
range of the interaction has to be short compared to the size of
the trap and is then the solution of an exactly solvable model. The 
pointlike interaction implies that the relative angular momentum $l=0$.
The formula has been extended to $l>0$ assuming  that the range of the
interaction is still small. \cite{suz09}

In the present work, restricted to $l=0$, the expression for the
two-particle energy spectrum as derived in ref. \cite{bus98} 
is  used to  find the \it shifts \rm in 
energy-levels from the non-interacting to the interacting spectrum.
It is shown that these shifts converge to the phase-shift approximation
for large H.O. quantum-numbers $n$ and for small as well as large 
scattering lengths.
Calculations, shown in Sect. 2,  moreover show that the
phase-shift approximation and the 'exact' result for the two-body
spectrum  obtained  by  
Busch et al. give practically indistinguishable results.

The good agreemnet with the phase-shift approximation for the two-body
spectrum may be unique to the H.O. trap.

For comparison, in the similar case of two particles interacting in 
a spherical square well the calculations presented
in Sect. 3 shows that  the phase-shift approximation
is viable  only in the limit of small (in units of box-radius)
scattering lengths  and large $n$-values and  it is in general a
much worse an approximation  than it is for the H.O.

\subsection{Spherical square-well}
If the trap is a spherical square-well (or box) the shift in energy 
spectrum due to the two particle's mutual interaction can, as shown
below in Sect. 3, easily be obtained without reference to a potential 
model.\cite{fu55,got} In the limit of a large size of the box this shift
is $\propto \delta(k)$, the scattering phase-shift at momentum $k$.
(Only  angular momentum $l=0$ is considered here although most results
are also valid for all values of $l$.)

A more elaborate but instructive method of obtaining the energy-shift 
was used in  early attempts of developping a many-body theory of nuclei.
Brueckner initially assumed the in-medium (effective) two-body 
interaction to be $\propto \tan\delta(k)$.  \cite{bru54}
The initial problem studied was that of an 'infinite' system of nuclear
matter for the purpose of calculating binding energy and saturation
properties of this system. The $\tan\delta(k)$ approximation came from 
the assumption that the Reactance matrix \cite{newton},
could be used as an approximation for the in-medium interaction in 
the many-body problem. (The diagonal part of this matrix is 
$\propto \tan\delta(k)$). The definition of the Reactance
matrix involves a principal value integration over a \it continuous \rm
spectrum. It was assumed that the numerical summation over the 
closely spaced level-spectrum of nucleons in a very large ('infinite') 
box would give an equivalent result.
This assumption was substantiated  by  Reifman and DeWitt\cite{rei56}.
It was however soon realized to be incorrect. Several authors (one of them
DeWitt) showed that the correct limit for two particles in a big box would 
yield an in-medium interaction $\propto \delta(k)$ rather than
$\tan\delta(k)$.  \cite{fuk56,dew56,rie56}
\footnote{For small values of $\delta$ and  at low density this might 
not make any difference but with $\delta\approx \pi/2$ (large scattering 
length), a situation often discussed in recent works, it obviously
does.}

The Reactance matrix is part of scattering theory. The nuclear matter
problem assumes a box, although large but still finite  with boundary 
conditions different from that of scattering theory. 
The proofs in the  papers refered to above differ, but the essential point 
is that
the integration over the continuous spectrum vs the summation over 
the discrete spectrum of particles in a box differ, even though the 
level-spacing in a big box goes towards zero as the box-size increases. 
The principal value integration relates to the scattering  problem  with
boundary conditions different from that for a box where the wave-functions
are zero at the edge of the box even in the limit of the box being
'infinitely' large. 

The exact statement of the results in the referenced papers is that the
energy-shift $\Delta$ due to the interaction of two-particles confined 
in a spherical box in the limit when the size of the box approaches 
infinity is given by 
\begin{equation}
\Delta=S\times dE
\label{DS}
\end{equation}
where $dE$ is the spacing between the levels of the unperturbed spectrum
and
\footnote{It was pointed out by DeWitt\cite{dew56}, after a comment by 
Brueckner that the result may be different for a box other than spherical. }
\begin{equation}
S=-\frac{\delta(k)}{\pi}.
\label{d}
\end{equation}
 
This energy shift implies that the \it diagonal \rm part of the 
'effective' interaction in a big box is given by
\begin{equation}
V_{\it eff \rm}(k)=4\pi \frac{\delta(k)}{k}
\label{veff}
\end{equation}

In nuclear matter studies (implying an infinite box) this has been
referred to as the 'phase-shift approximation' a term adopted here.  
It was for example used in early calculations on the 
neutron-gas\cite{bru60,soo60}.  Many-body calculations 
do in general also require \it off-diagonal \rm elements of the 
'interaction, as in the calculation of the Brueckner reaction-matrix. 
In some early nuclear matter studies, these were obtained by assuming
$V_{\it eff \rm}$ to be a separable interaction with the diagonal given
by eq. (\ref{veff}).\cite{hsk83}.
 
It is of interest to compare the energy-shifts in a \it finite \rm
size box with those of the H.O. trap. Results of these calculations are
shown in Sect. 3.

\section{3-D Harmonic Oscillator Well}
Units of energy and length are here chosen to be $\hbar \omega$ and 
$a_{o}=\sqrt{\hbar/m\omega}$ respectively.
The total energy for two non-interacting particles with zero angular
momentum in a 3-D H. O. well is then
$E_{tot}=E_0+E_{cm}=2n+3$ ($n=0,1,2...$). Assuming the center=of-mass
energy to be  $E_{cm}=\frac{3}{2}$ the energy of relative motion is
$E_0=2n+\frac{3}{2}$.\footnote{Only the dependence of the quatum-number
of relative motion is relevant for the following}

With the particles interacting, $E_0\rightarrow E$ and with
$\eta=2E$  the energy spectrum is obtained from  \cite{bus98,jon02}
\begin{equation}
\tan\delta(k)=-\frac{\sqrt{\eta}\Gamma[(1-\eta)/4]}{2\Gamma[(3-\eta)/4]}
\label{busch}
\end{equation}
where $k=\sqrt{\frac{\eta}{2}}/a_{o}$.

This expression for the energy-spectrum of two particles trapped in a
H.O. well has already played an important role in interpreting
experimental data in atomic physics.  (E.g. refs.\cite{sto06,osp06})
It is cited in numerous recent articles related to nuclear
structure.(E.g. ref. \cite{rot13,yan13}).
It has been extended to relative angular momenta $l>0$ by Suzuki et al
\cite{suz09}.

It is here recast in a different form to explicitly 
show the energy-\it shifts \rm $\Delta=E-E_0$ incurred by the 
two-body interaction.  One finds $$\eta=4n+3+2\Delta$$.
After substitution in eq. (\ref{busch}) and 
using the reflection formula for $\Gamma$-functions \cite{abr72}
\begin{equation}
\Gamma(x)=\frac{\pi}{\Gamma(1-x)\sin(\pi x)}
\label{refl}
\end{equation}
one finds

\begin{equation}
\tan \delta(k)=-A(z)\tan(\frac{\Delta}{2\pi})
\label{busch2}
\end{equation}
or 
\begin{equation}
\Delta=-\frac{2}{\pi}arctan\frac{tan(\delta(k))}{A(z)}
\label{bush2a}
\end{equation}
where

\begin{equation}
A(z)=\frac{\sqrt{z-\frac{1}{4}} \Gamma(z)}{\Gamma(z+\frac{1}{2})}
\label{busch3}
\end{equation}
with
\begin{equation}
z=n+\Delta/2+1.
\label {z}
\end{equation}

Eq. (\ref{bush2a}) is equivalent to eq. (\ref{busch}), first derived by
Busch et al \cite{bus98}  although here rewritten 
in terms of the shift $\Delta$ instead of the energy $E=\eta/2$.
Of particular interest here is the function $A(z)$.

\begin{figure}
\centerline{
\psfig{figure=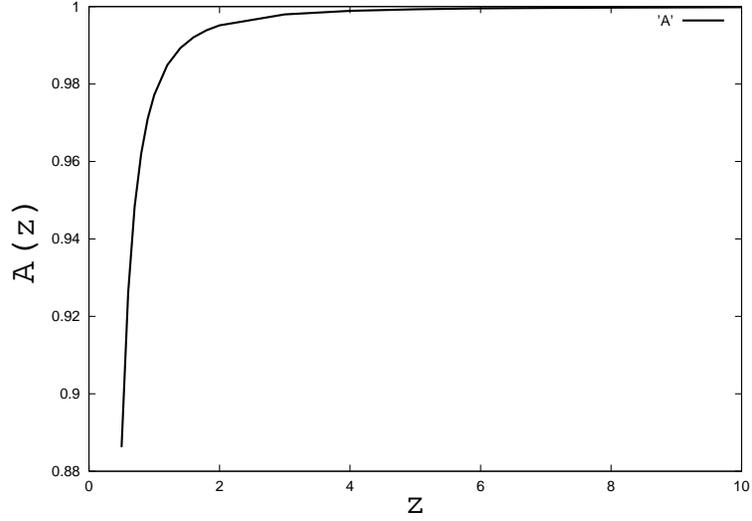,width=7cm,angle=-90}
}
\vspace{.0in}
\caption{
Function $A(z)$ defined in the text. Notice the asymptotic convergence
of $A(z)\rightarrow 1$.
}
\label{bush1}
\end{figure}

Fig. \ref{bush1} shows  a rapid convergence 
of this function to its  
asymptotic value, i.e. $A(z)\rightarrow 1$ for increasing  values of $z$,
and in this limit  eq. (\ref{bush2a})  reduces to eq.(\ref{d}), 
referred to as the phase-shift-approximation. 
\footnote{Using the asymptotic formula for
the $\Gamma$-function in ref.\cite{abr72} one finds $A(z)\rightarrow
\sqrt\frac{z-\frac{1}{4}}{z}$, indicating an appreciably slower
convergence to the asymptotic value than the exact calculation that is
shown in Fig. \ref{bush1}.}

This also occurs, irrespective of the the value of $A(z)$, when
$\delta(k)=0$ and more importantly
when $\delta(k)=\pm \frac{\pi}{2}$ in which case $\Delta=\mp 1$, a
well-known result.\cite{bus98,jon02}.

A correction $d\Delta$ to the phase-shift approximation of $\Delta$ (i.e.
for values of $z$ when $A(z)\leq 1$) is obtained from
\begin{equation}
d\Delta\approx
-\frac{1}{\pi}(1-A^{-1}(z))sin(2\delta)
\label{correct}
\end{equation}
It is damped with decreasing values of
$\sin{2\delta}$  and goes to zero when $\delta=\pm\frac{\pi}{2}$  (and
when $\delta=0$) in accordance with above.

\begin{figure}
\centerline{
\psfig{figure=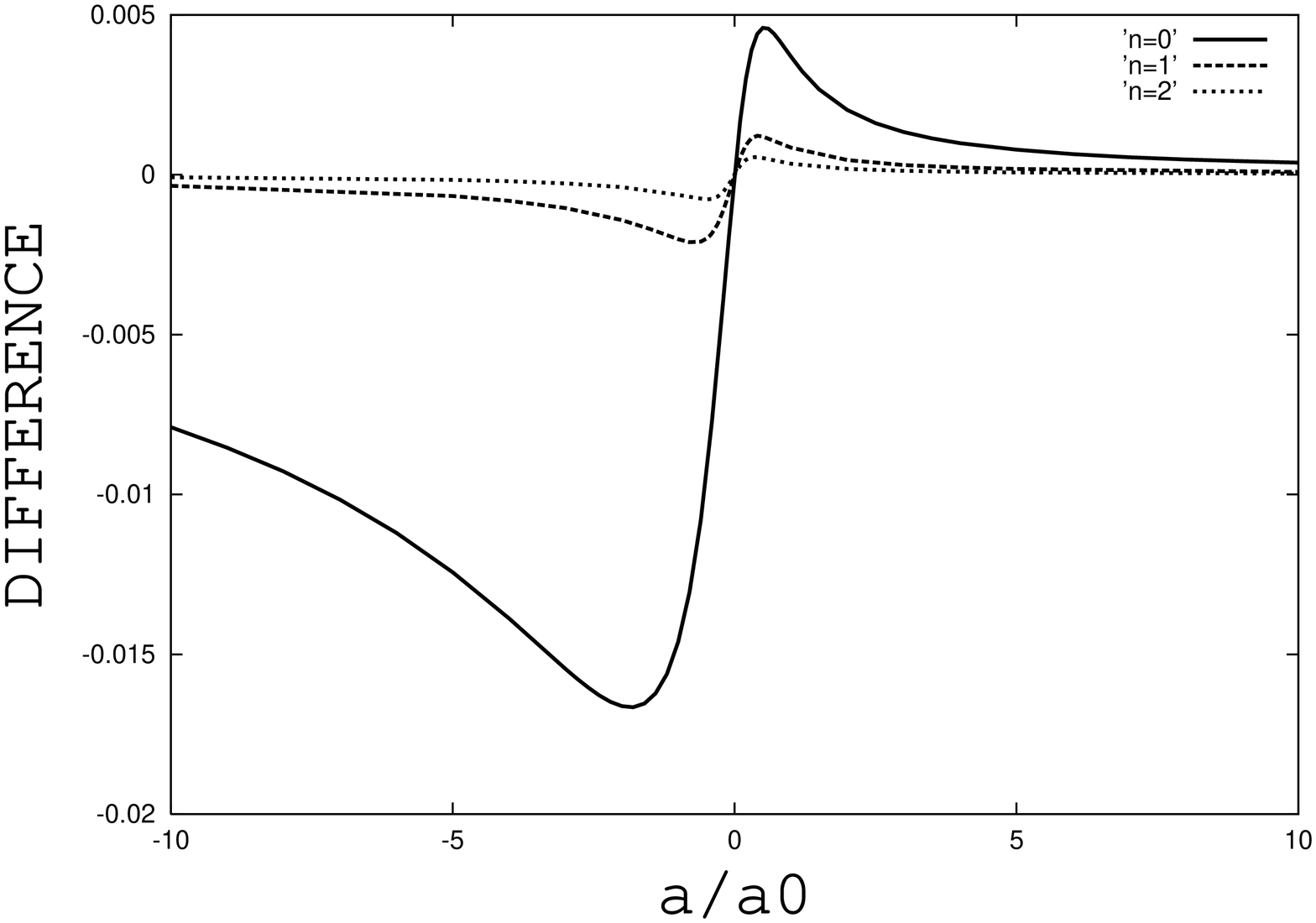,width=7cm,angle=-90}
\psfig{figure=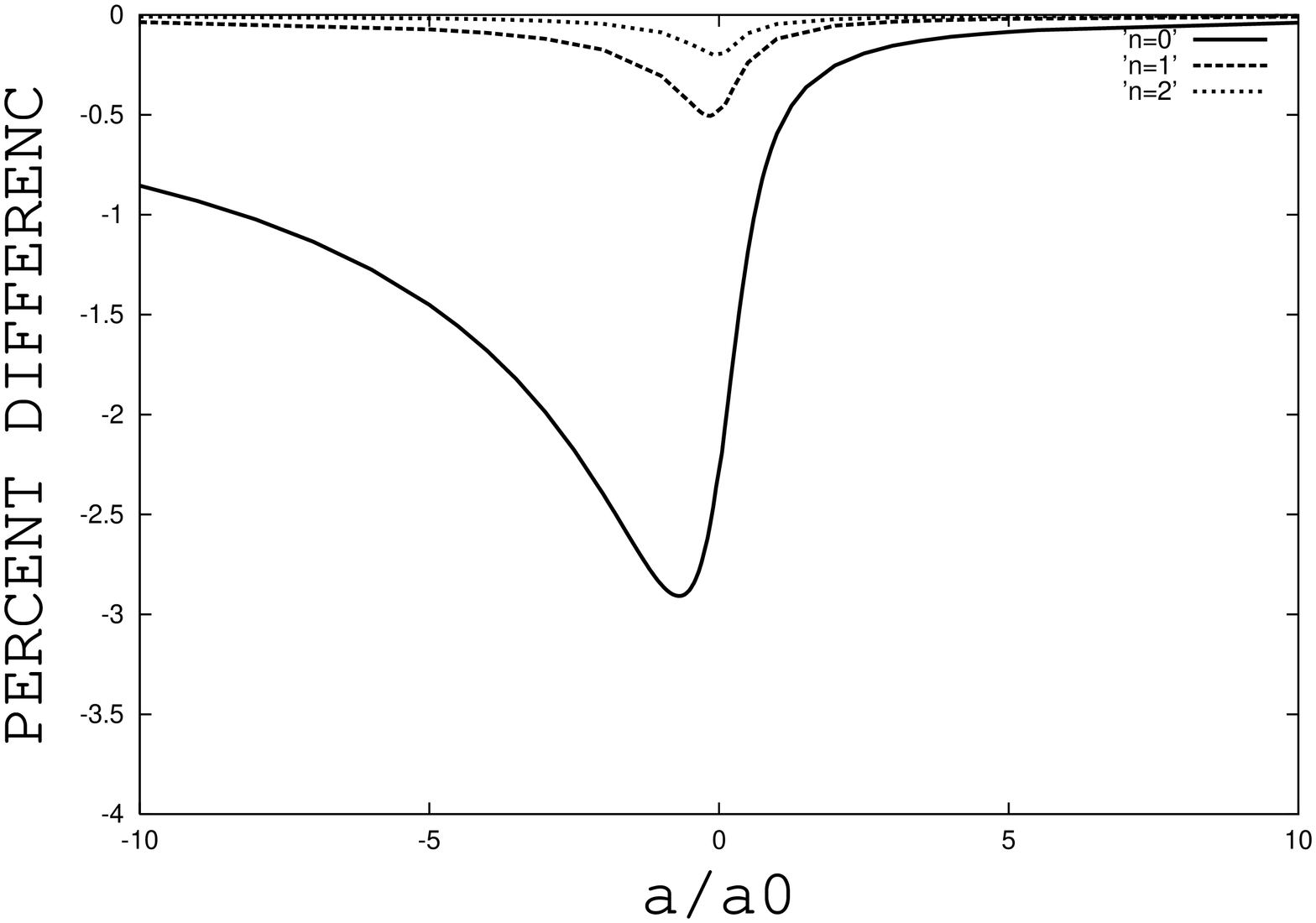,width=7cm,angle=-90}
}
\vspace{.0in}
\caption{
Curves to the left show  the differences between energy-shifts calculated
exactly, eq. (\ref{bush2a}) and by the phase-shift approximation 
(\ref{d}) for two-particles in a H.O. trap  for  quantum-numbers
$n=0,1$ and $2$, with $n=0$ showing the largest difference.
To the right is shown the same differences but expressed in percent.
A comparison  with the similar plot in Fig. \ref{bush3} for the
square well trap shows  that the convergence with $n$ is faster in this
case.}
\label{bush2}
\end{figure}

The left part of Fig. \ref{bush2} shows the difference between the 
energy shifts calculated by  eq. (\ref{bush2a}), and the
phaseshift-approximation, eq. (\ref{d}), respectively for H.O. 
Quantum-numbers $n=0,1$ and $2$.  Results are shown  as a function 
of scattering length $a$ in units of $a_{o}$. The difference (error)  
is seen to decrease rapidly with increasing $n$ and  (as also shown above) 
with increasing value of $|a|$. The peaks are a result of the competition
between the $A(z)$ and $\sin{2\delta}$ terms respectively shown in eq.
(\ref{correct}). 
The right part of the Fig. \ref{bush2} shows the same differences but 
in $\%$ of the exact result again showing the rapid decrease with 
increasing values of $n$ and with $|a|$. 
One finds  larger differences  for $n=0$ , while  appreciably
smaller for $n=1$ and $n=2$ . This is of course related to the function
$A(z)$. The smallest value of
this  function  is for for the smallest value of $z=\frac{1}{2}$ 
(see Fig. \ref{bush1}), which occurs for $n=0$ $S=-\frac{1}{2}$, $n=1$
$S=-\frac{3}{2}$  etc.

\begin{figure}
\psfig{figure=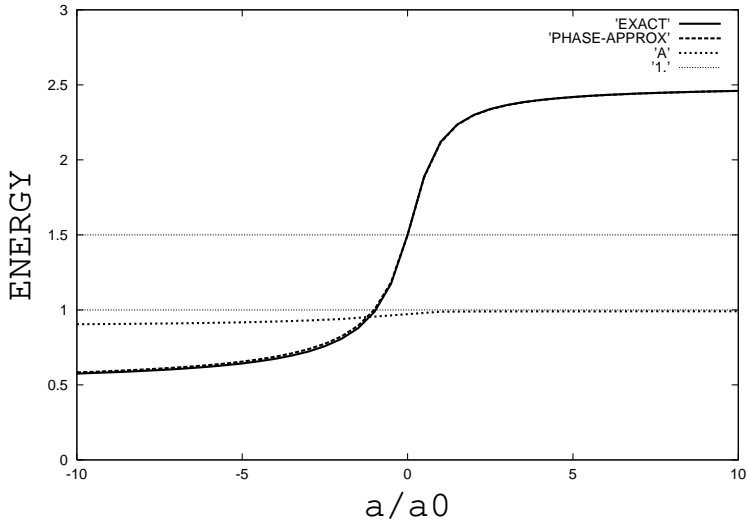,width=7cm,angle=-90}
\vspace{.0in}
\caption{The solid line(s) show(s) the exact (eq. (\ref{busch}) 
two-body energy as a function of scattering length and also from the 
phase-shift approximation.  The two lines practically coincide. 
A slight difference may be seen for negative scattering lengths. 
The comparison is for the worst scenario, the lowest H.O. energy, 
i.e. $n=0$.  Compare with Fig. \ref{bush2} that  shows the difference 
between the two curves.  The broken line shows the function $A(z)$, 
eq. (\ref{bush1}), where $z$ is obtained from eq. (\ref{z}). 
It is seen to converge towards $A(z)=1$ for positive scattering
lengths. The non-intercting H.O. energy is here $n+\frac{3}{2}=1.5$.
}
\label{bush4}
\end{figure}

The largest discrepancy is consequently  expected to be found for  $n=0$ 
and for $a<0$, where the shift is negative. The comparison between the
exact result given by eq. (\ref{busch} and the phase-shift approximation 
for this worst case is shown in
Fig. \ref{bush4}. It shows practically complete overlap between the
two curves even for this worst case. The 'correcting' factor $A(z)$ is also shown in this figure.
It is seen that $A(z)\approx 0.9$ for negative scattering lengths but
approaches the value $A(z)=1$ for positive scattering lengths which is
consistent with eq. (\ref{z}).
For comparison, calculations show that $A(z)\leq 0.991$ when $n=1$ and
with complete overlap between exact and the approximate results.

The conclusion is that the phase-shift approximation for the effective
interaction applies equally well or even better in the H.O. case 
than it does in the square-well case shown below in Sect. 3.
Again, of course, in a many body environment (three or more particles) 
also off-diagonal elements are needed. J. Rotureau \cite{rot13} solves 
this problem by generating an interaction by a unitary transformation 
of the H.O. two-particle spectrum given by eq.  (\ref{busch}).

A main result of this investigation is that the energy spectrum of 
two particles interacting in a H.O. trap can 
be simply constructed from the free-scattering phase-shifts  as
follows:

\begin{equation}
E(a)=E_0-2\frac{\delta(k)}{\pi}
\label{fin}
\end{equation}
where $E$ and $E_0$ are the perturbed and un-perturbed levels
respectively with $k=\sqrt{E(a)}$ and $tan(\delta(k))=-ak$. 

As written, this result refers to two particles with point-like
interactions and angular momentum $l=0$.  It can be extended to 
$l>0$ but is then expected to be less accurate, as is also the case 
with the original formula eq. (\ref{busch}).

\section{Spherical Square Well}
It is of interest to compare the above result obtained with a H.O.
trap with that of two particles in a spherical box of finite size
as this may better correspond to some experimental setups.

The problem of two particles  interacting in a  box of \it infinite \rm 
size was considered by several authors 
\cite{fu55,got,fuk56,dew56,rie56} referred to in the Introduction.  
with the result given by eq.  (\ref{d}), the phase-shift approximation.
The special interest was, at the time, 
that of nuclear matter, a system of nucleons in a large but finite 
enclosure or of free electrons in a metal.

It is of interest to compare the result obtained with a H.O.
trap with that of two particles in a spherical box of \it finite \rm
size.
This may better correspond to some experimental setups.

The simplest solution \cite{fu55,got} of the problem at hand is to 
explicitly consider the wave-functions of the two particles 
in this box. It is an example of an exactly solvable model in
quantum mechanics.\cite{mahan}
Considering only $s$-states, the radial wavefunctions of free 
non-interacting particles are (easily extended to angular momentum $l>0$)
\begin{equation}
\Psi(r) \propto \frac{1}{r}\sin(k^{(0)}r)
\label{ucorr}
\end{equation}
With the two particles interacting, the wave-function outside 
the range of the interaction (assumed $\ll R$ or 'pointlike') is:

\begin{equation}
\Psi(r)\propto \frac{1}{r}\sin(kr+\delta(k))
\label{corr}
\end{equation}

The boundary condition implies that  wave-functions vanish at the
boundary of the sphere.
With units of energy and length  chosen to be $\hbar^2/{2m}$ and
well-radius $R$ respectively one finds for the non-interacting case 
\begin{equation}
k^{(0)}_n=n\pi
\label{ucorr1}
\end{equation}
and for the interacting case
\begin{equation}
k_n+\delta(k_n)=n\pi
\label{corr1}
\end{equation}

The spacing between unperturbed levels is
$$dE=2n\pi^2$$

The energy shift is
$$\Delta=k_n^2-{k^{(0)}_n}^2=-\delta(k_n)(k_n+k_n^{(0)})$$
so that by eq. (\ref{DS}) 

\begin{equation}
S=-\left(1-\frac{\delta(k_n)}{2n\pi}\right)\frac{\delta(k_n)}{\pi}
\label{sphe1}
\end{equation}
showing that for large $n$ and/or small $\delta$ 
$$S\rightarrow -\frac{\delta(k_n)}{\pi}$$ which is  the phase-shift
approximation  (\ref{d}).

\begin{figure}
\centerline{
\psfig{figure=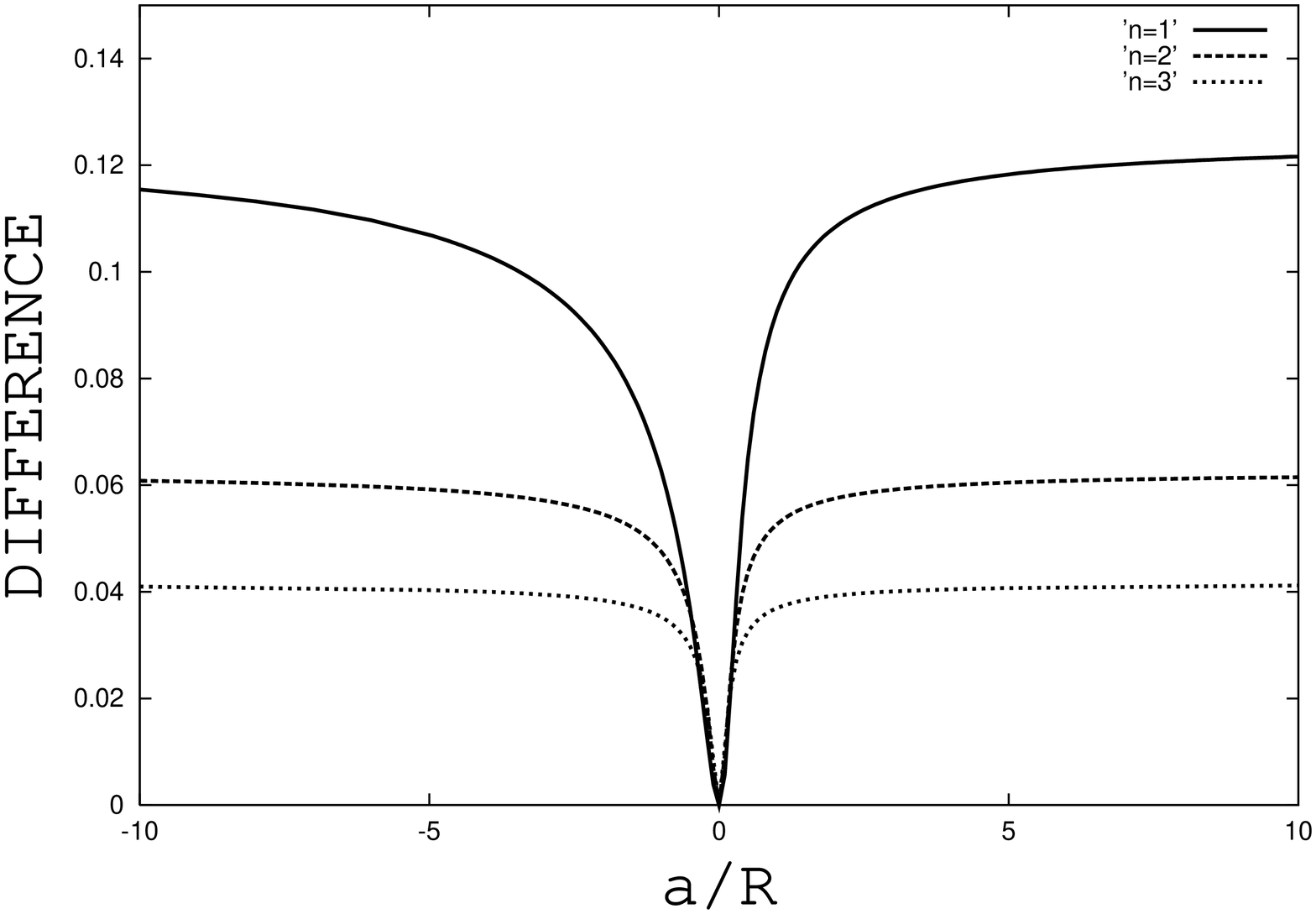,width=7cm,angle=-90}
\psfig{figure=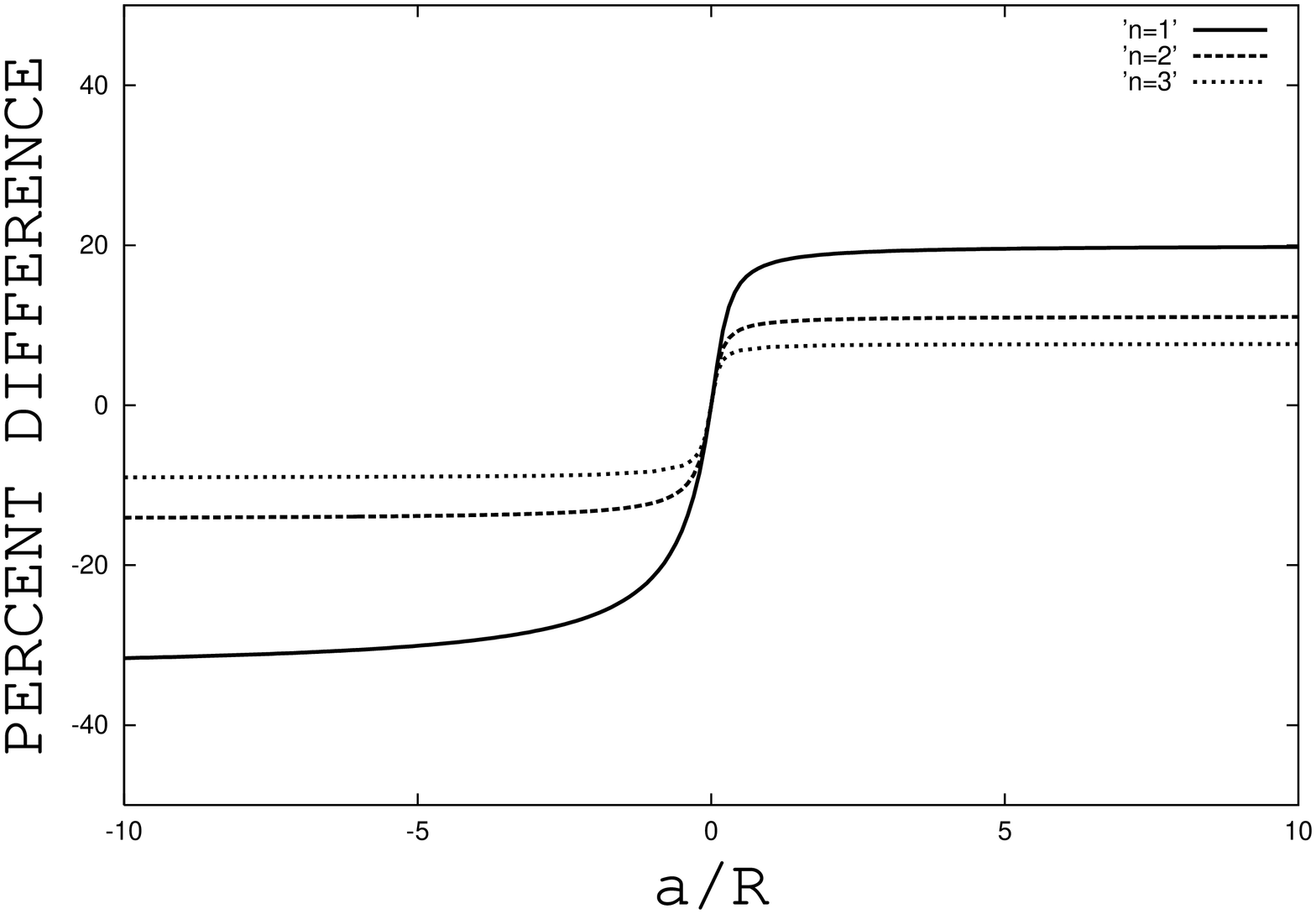,width=7cm,angle=-90}
}
\vspace{.0in}
\caption{
Curves to the left show  corrections to the
phase-shift approximation 
for two-particles in a spherical square well trap  for  quantum-numbers
$n=1,2$ and $3$, with $n=1$ showing the largest difference.
To the right is shown the same differences but expressed in percent.
}
\label{bush3}
\end{figure}

Fig. \ref{bush3} shows  the corrections 
$\frac{\delta(k_n)^2}{2n\pi^2}$
to the phase-shift approximation as a function of 
scattering length $a$ and for some indicated values of quantum-number $n$.

The phaseshift is here a function of $a$ and $k_n$ by eq. (\ref{corr1})
which is solved selfconsistently.

\begin{figure}
\psfig{figure=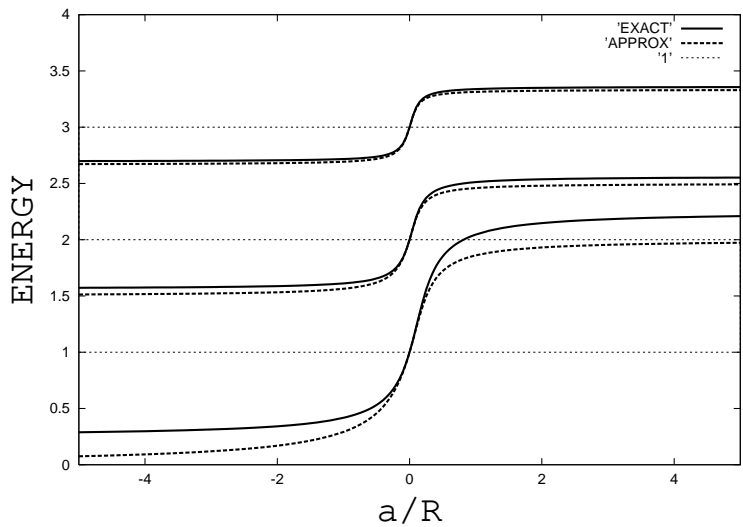,width=7cm,angle=-90}
\vspace{.0in}
\caption{
The solid lines show the exact two-body spectra for $n=1,2$ and $3$, while
the broken lines show the result of the phase-shift approximation. 
The energies are normalised to the uncorrelated enrgies.
}
\label{bush10}
\end{figure}

A comparison with the analogous result for the H.O. in Fig.
\ref{bush2}  shows notable
differences. While the  largest deviation from the phase-shift
approximation in the H.O. case
is $3\%$ (and typically much smaller) , it is in the
square well case seen to be  as large as $30\%$. 
In the H.O. the  largest error is for small scattering lengths
while the error goes to zero for large scattering lengths. This  is opposite 
to the situation encountered for the  square well. 
Fig. \ref{bush3} does  however show the known
result that $S\rightarrow -\frac{\delta(k)}{\pi}$ 
when $a/R \rightarrow 0$
\cite{fu55,got,fuk56,dew56,rie56}.

Fig. \ref{bush10} shows the two-body spectra calculated exactly compared
with those obtained in the phase-shift approximation. The agreement
between the the two sets of curves  increases with increasing $n$ as 
well as increasing $R$, i.e. size of the box but a comparison with 
Fig. \ref{bush4} shows that the approximation works better in the H.O.
case for the low-lying states. 

\section{Summary}
The validity of the phase-shift approximation,
$\Delta=-\frac{1}{\pi}\delta(k)dE$ for the energy-shift 
due to the interaction of two partcles in a trap, where $\delta(k)$ is the
free particle scattering phase-shift and  dE  the energy-level spacing 
of the unperturbed spectrum,  has been  investigated. 

The traps considered were a 3-D H.O. and a spherical 
infinite square well (box).  The square-well  case was treated a long time ago
as referenced above, with the main emphasis on the large size limit of the 
box with the result expressed by eq. (\ref{d}), the phase-shift
approximation. 
Somewhat of a surprise was the finding 
that a not only  somewhat similar situation exists for the H.O. trap but that 
the approximation is so much more accurate for this trap in the region
of trap parameters of interest for finite systems.  
Eq.(\ref{fin}) can therefore  be considered as an
alternative to the formula of Busch et al \cite{bus98}.
In addition to being simple to use for calculating the two-body spectra
in the H.O. it can also very simply be used to extract  phase-shifts
from experimental data.

The differences between the results
for the two respective traps are seen by comparing Fig. \ref{bush2}  
for the H.O.trap with Fig. \ref{bush3} 
for the square-well trap.  While in the  square well
case, the differences (errors) by using the phase-shift approximation are as
large as $30\%$, the differences are much smaller and even practically zero
in the H.O. case.   The accuracy of the  approximation in the
H.O. case is also seen in Fig. \ref{bush4} with curves practically
overlapping  even in the worst case, $n=0$.
For comparison, the scenario is rather different in the square well case shown 
by Fig. \ref{bush10} where for the low-lying states the approximation is 
only valid  for small values of $a/R$.

Another difference between the H.O. and the box rsults is that, 
with $a\rightarrow \pm \infty$ the
phase-shift approximation becomes perfect in the H.O. case while this is
not so in  case of the box. In both cases the approximation 
becomes good for small values of the scattering-lengths, which is not
surprising as the Born approximation would then be applicable.
 
The results presented here were for zero-range potentials, 
i.e. for zero effctive range i.e. with $tan\delta(k)=-ka$.
Jonsell \cite{jon02} found the finite range-corrections to be small 
in the H.O. case. The 'small' range is of course in relation to the
physical size of the system, i.e. the length parameters defined above.
This subject  was also the subject of a recent publication.\cite{luu10}

This investigation can be extended to angular momenta $l>0$.

The question arises of course whether the phase-shift approximation has a more
general validity e.g. for two particles in a \it non-spherical \rm atomic
nucleus or  for particles in a cubical box.
It was however already illustrated above  that the situation is
rather different in the 3-D H.O. case compared to the case of a spherical
well. The conceptual differences  between scatterings and interactions in
the spherical vs cubical box was already pointed out by DeWitt in a 
detailed discussion of this matter.
Quote\cite{dew56}:\it Only the spherical box (with spherical waves) is suitable
for establishing a connection between single scattering processes and
discrete spectrum theory....\rm
but the problem of two atoms in a anisotropic H.O. trap was considered in ref.
\cite{idz05}.
A related problem that of the relation between the energy spectrum and
scattering phases of  two particles in a 
cubical box but with \it periodic \rm boundary conditions 
shift was however solved by L\"uscher\cite{lu91} . 
but he problem of two atoms in anisotropic H.O. trap was considered in ref.
\cite{idz05}.

\end{document}